\documentclass{article}
\usepackage{ijcai24}

\usepackage{times}
\usepackage{soul}
\usepackage{url}
\usepackage[bookmarks=false]{hyperref}
\usepackage[hyphenbreaks]{breakurl}
\usepackage[utf8]{inputenc}
\usepackage[small]{caption}
\usepackage{graphicx}
\usepackage{amsmath}
\usepackage{amsthm}
\usepackage{booktabs}
\usepackage{algorithm}
\usepackage{algorithmic}
\usepackage[switch]{lineno}

\usepackage{xcolor}
\usepackage{comment}
\usepackage{wrapfig}

\title{VRPD-DT: Vehicle Routing Problem with Drones Under Dynamically Changing Traffic Conditions}

\author{
    Navid Imran and Myounggyu Won
    \affiliations
    Unviersity of Memphis, TN, United States
    \emails
    \{nimran, mwon\}@memphis.edu
}

\begin{document}

\maketitle

\begin{abstract}
The vehicle routing problem with drones (VRP-D) is to determine the optimal routes of trucks and drones such that the total operational cost is minimized in a scenario where the trucks work in tandem with the drones to deliver parcels to customers. While various heuristic algorithms have been developed to address the problem, existing solutions are built based on simplistic cost models, overlooking the temporal dynamics of the costs, which fluctuate depending on the dynamically changing traffic conditions. In this paper, we present a novel problem called the vehicle routing problem with drones under dynamically changing traffic conditions (VRPD-DT) to address the limitation of existing VRP-D solutions. We design a novel cost model that factors in the actual travel distance and projected travel time, computed using a machine learning-driven travel time prediction algorithm. A variable neighborhood descent (VND) algorithm is developed to find the optimal truck-drone routes under the dynamics of traffic conditions through incorporation of the travel time prediction model. A simulation study was performed to evaluate the performance compared with a state-of-the-art VRP-D heuristic solution. The results demonstrate that the proposed algorithm outperforms the state-of-the-art algorithm in various delivery scenarios.
\end{abstract}

\section{Introduction}

The past decade has witnessed explosive growth in the use of unmanned aerial vehicles (UAVs) or drones \cite{otto2018optimization}. Drones have been integrated into a diverse range of applications, including smart farming \cite{tripicchio2015towards}, disaster response \cite{rashid2020socialdrone}, and remote sensing \cite{inoue2020satellite}. This paper focuses specifically on drone-based last-mile delivery, an emerging technology poised to revolutionize traditional parcel delivery systems. Drones offer unparalleled speed and accessibility, as they are not bound by road infrastructure, making them a promising solution for efficient and effective deliveries. The collaboration between drones and trucks, where trucks work in tandem with drones to serve customers, can significantly enhance the efficiency of parcel delivery. Numerous tech companies such as Amazon \cite{amazon}, Alibaba \cite{alibaba}, and Google \cite{google} as well as logistics companies like FedEX \cite{fedex_2022}, DHL \cite{dhl}, and UPS \cite{intelligence} started developing and deploying drone-based last-mile delivery solutions. 

The literature on drone-based last-mile delivery has primarily concentrated on determining the optimal routes of trucks and drones to minimize the total operational cost. A pioneering work conducted by Murray and Chu~\cite{murray2015flying} introduced two main last-mile delivery problems called the Flying Sidekick Traveling Salesman Problem (FSTSP) and the Parallel Drone Scheduling Traveling Salesman Problem (PDSTSP). In FSTSP, a drone is carried on a truck and deployed to serve a customer while the truck proceeds to serve another customer. After completing its task, the drone returns to the truck. Conversely, in PDSTSP, trucks and drones independently serve customers starting from the depot. Two primary research streams have emerged from FSTSP, namely the Traveling Salesman Problem with Drone (TSP-D)~\cite{vasquez2021exact,roberti2021exact} and the vehicle routing problem with drones (VRP-D)~\cite{wang2017vehicle,poikonen2017vehicle}. While a single truck with a single drone is assumed in TSP-D, multiple trucks and multiple drones are used in VRP-D. In this paper, our primary focus is on VRP-D given its applicability to a wider array of delivery scenarios. An in-depth review of the various heuristic solutions proposed to address VRP-D is discussed in Section~\ref{sec:related_work}.

Despite the effectiveness of state-of-the-art heuristic algorithms in solving VRP-D and its variants, a significant drawback of current solutions is their lack of consideration for the impact of dynamically fluctuating traffic conditions. This omission could result in suboptimal solutions in real-world applications, as a solution deemed optimal at a particular time may no longer be optimal as traffic conditions change over time. To address the limitation of existing solutions, this paper presents a novel problem, referred to as the Vehicle Routing Problem with Drones under Dynamically Changing Traffic Conditions (VRPD-DT). The primary objective is to determine the optimal routes for trucks and drones that minimize the overall operational cost while considering dynamically changing traffic conditions. To address VRPD-DT, we introduce a heuristic solution based on machine learning-driven travel time prediction that estimates the time taken between two locations that a truck is planned to visit. More specifically, we develop a novel cost model utilizing the actual traveled distance and the predicted travel time. A Variable Neighborhood Descent (VND)-based heuristic solution is then designed, integrating the cost model. We train the travel time prediction model using the New York taxi dataset~\cite{NYC-tlc} containing over 21.9 million taxi trips and conduct a computational study to evaluate the performance of our solution. The results demonstrate that the proposed solution outperforms the state-of-the-art VRP-D algorithm~\cite{kuo2022vehicle} in various delivery scenarios. The following is a summary of our contributions.

\begin{itemize}
	\item We present the first-of-its-kind vehicle routing problem with drones under dynamically changing traffic conditions (VRPD-DT).
	\item We introduce a novel cost model driven by machine learning, leveraging real-world traveled distance and predicted travel time between two points that a truck is scheduled to visit. This model enables accurate estimation of the operational costs associated with the routes, enhancing the optimization process.
	\item A novel variable neighborhood descent (VND)-based heuristic solution is proposed to solve VRPD-DT. 
	\item An extensive simulation study is conducted to demonstrate the significant performance gain of the proposed heuristic solution compared with a state-of-the-art approach in solving VRP-D under dynamically changing traffic conditions. 
\end{itemize}

This paper is structured as follows: Section~\ref{sec:related_work} begins with a review of the latest literature on VRP-D solutions. This is followed by the preliminaries in Section~\ref{sec:vrptwd}. In Section~\ref{sec:motivation}, we conduct a motivational study to highlight the limitations of existing approaches. Section~\ref{sec:proposed_system} presents the details of our proposed heuristic solution. Simulation results are discussed in Section~\ref{sec:results}, and we conclude in Section~\ref{sec:conclusion}.

\section{Related Work}
\label{sec:related_work}

In this section, we review the literature on drone-based last-mile delivery focusing on VRP-D. VRP-D is to find the optimal routes of trucks and drones~\cite{wang2017vehicle}. More specifically, in this problem, drones are launched from a truck at any location including the customer locations and the depot to serve customers. After serving customers, the drone returns to the truck at a rendezvous location. There are various kinds of heuristic solutions developed to solve VRP-D. Poikonen \emph{et al.}, in particular, considered the constrained battery of drones as well as different distance metrics~\cite{poikonen2017vehicle}. Sacramento \emph{et al.} developed an adaptive large neighborhood search metaheuristic solution by considering the time-limit constraint~\cite{sacramento2019adaptive}. Kitjacharoenchai \emph{et al.} considered the constraints related to the drone launch and delivery time~\cite{kitjacharoenchai2019multiple}. Murray \emph{et al.} developed a solution taking into account heterogeneous drones ~\cite{murray2020multiple}. 

Kuo \emph{et al.} considered the customer time windows in solving VRP-D~\cite{kuo2022vehicle}. In their extended version of VRP-D, a time window is associated with a customer, which means that the customer must be visited either by a drone or a truck within the time window. To solve this problem, they presented a heuristic based on a variable neighborhood search. Wang \emph{et al.} defined an extension of VRP-D called the truck–drone hybrid routing problem with time-dependent road travel time drones (TDHRP-TDRTT)~\cite{wang2022truck}. A difference compared to the traditional VRP-D is that they considered traffic conditions to find the optimal routes for trucks and drones. To solve the problem, the authors develop an iterative local search algorithm. 

A limitation of existing VRP-D solutions is that they rely on a simplifying assumption that a drone serves only one customer per trip~\cite{gu2022hierarchical}. To address this limitation and enhance the practicality, Gu \emph{et al.}~\cite{gu2022hierarchical} developed a solution that allows a drone to serve multiple customers per trip. Huang~\emph{et al.} proposed a heuristic solution based on an ant colony optimization (ACO)~\cite{huang2022solving}. Nguyen \emph{et al.} introduced an extension of VRP-D with additional constraints~\cite{nguyen2022min}. More specifically, the capacity of a truck in terms of the total weight of parcels was considered, and the total operation time of both trucks and drones was taken into account to ensure that trucks and drones are used only for the duration of a pre-defined value. 

Rave \emph{et al.} introduced a new notion to VRP-D which is called the micro depots~\cite{rave2022drone}. In their problem, drones can be launched either from trucks or from these micro depots. Considering the micro depots, the authors developed a heuristic solution based on an adaptive large neighborhood search. Montana \emph{et al.} gave a special emphasis on the environmental effect of solutions for VRP-D~\cite{montana2022novel}. Specifically, they performed an analysis on the impact of drone-based last-mile delivery on sustainability in terms of carbon emission. Sitek \emph{et al.} defined the problem called the extended vehicle routing problem with drones (EVRP-D) which is different from the traditional VRP-D in that they considered mobile points called as mobile hubs (similar to the micro depots~\cite{rave2022drone}) where drones can be launched~\cite{sitek2022proactive}. They developed a genetic algorithm to minimize the operational cost and selection of mobile hubs. Wu \emph{et al.} concentrated on the effect of the payload and flight time on the energy consumption of drones and develop a heuristic algorithm based on variable neighborhood descent algorithm~\cite{wu2022collaborative}.

While state-of-the-art heuristic algorithms for VRP-D and its variants are highly effective, they often overlook a critical aspect: the impact of dynamically fluctuating traffic conditions. This significant oversight in current methodologies can lead to solutions that quickly become outdated as the traffic landscape changes, potentially compromising efficiency and effectiveness. Our approach addresses this gap. We introduce a novel VRP-D solution that uniquely accounts for these dynamic traffic conditions. Our approach not only acknowledges the variable nature of real-world traffic but also adapts to it, ensuring that our solutions remain optimal over time.

\section{Preliminaries}
\label{sec:vrptwd}

This section presents the formal definition of VRP-D. Specifically, we focus on one of the latest variants of VRP-D, vehicle routing problem with drones that takes into account the customer time windows
(VRPTWD)~\cite{kuo2022vehicle}. The notations and assumptions introduced in this section are consistently used throughout the paper to explain our work.

In VRPTWD, a delivery scenario is modeled as a graph $G=(V,A)$. The set of nodes $V$ comprises customers and depots, \emph{i.e.,} $V=C \cup P$, where $C=\{1,...,n\}$ signifies a set of customers requiring service, while $P=\{0,n+1\}$ denotes the depots. Each customer $i \in C$ carries a demand $q_i$ in terms of the number of parcels to be delivered. The edge set $A$ consists of arcs between nodes. Each arc $(i,j) \in A$ is coupled with a distinct travel time for both a truck and a drone, represented by $t_{ij}$ and $t'_{ij}$ respectively.

The available delivery trucks and drones for delivery are denoted by $T=\{1, ..., n\}$ and $D=\{1, ..., n\}$, respectively. These trucks serve all customers in a manner that satisfies the demands $q_i$ of all customers $i \in C$. In particular, each truck has the load capacity $Qt$. Drones are loaded on a truck, and each drone has its own load capacity $Qd$. In VRPTWD, each customer $i \in C$ must receive service within a specified time frame $s_i$. This time frame is denoted as a time window $[o_i, e_i]$, where $o_i$ is the earliest and $e_i$ is the latest acceptable service time. We assume that if a vehicle arrives too early, the vehicle should wait until the time window for the customer opens. Additionally, $a_i^j$ denotes the arrival time of truck $j \in T$ at customer $i \in C$, and ${a'}_i^j$ is the arrival time of drone $j \in D$ at customer $i \in C$. If $a_i^j > e_i$ ($j \in T$, $i \in C$), or ${a'}_i^j > e_i$ ($j \in D$, $i \in C$), a penalty of $W(j)$ is imposed.

We define specific types of nodes within the set $C$ of customer nodes. First, we introduce the launching nodes, $C_L=\{0,1,\ldots,n\}$, a subset of $C$ where drones are launched. Additionally, we define the rendezvous nodes, $C_R=\{1,\ldots,n,n+1\}$, $C_R \subseteq C \cup P$, where trucks and drones meet. The subset of customer nodes that a drone can service, where the demand $q_i$ does not exceed the drone's capacity $Q_d$, is denoted by $C_D=\{0,1,\ldots,n\}$ and is also a subset of $C$. Considering a typical delivery scenario, a drone $v \in D$ is launched from a truck at a launching node $i \in C_L$, services a customer at node $j \in C_D$, and then returns to meet the truck at a rendezvous node $k \in C_R$. In this scenario, a binary variable $y_{ijk}^v$ takes a value of 1 if drone $v$ is deployed from node $i$, delivers to node $j$, and returns to node $k$. Moreover, $T_{\text{max}}$ represents the maximum duration of a delivery operation. The set $r_i^t \subseteq C$ denotes the customer nodes visited by truck $i \in T$, and $r_i^d \subseteq C$ signifies the nodes visited by drone $i \in D$."

After establishing all the necessary notations, the goal of VRPTWD is to optimize the routes for trucks and drones to minimize their combined operational costs. For trucks, the operational cost to traverse an arc $(i, j) \in A$ is calculated as $C_{ij} = (d_{ij} \cdot MC) \cdot FP \cdot FC$. Here, $d_{ij}$ represents the distance of the arc $(i, j) \in A$, $MC$ is the miles converter, $FP$ is the fuel price per liter, and $FC$ signifies the truck's fuel consumption rate. Additionally, the operational cost for a drone traveling the same arc is denoted by $C'_{ij} = \alpha \cdot C_{ij}$, where $\alpha$ is a multiplicative factor representing the relative cost efficiency of a drone compared to a truck.

\section{Motivational Study}
\label{sec:motivation}


\begin{wrapfigure}{r}{0.6\columnwidth}
	\begin{center}
		\vspace{-15pt}
		\includegraphics[width=\linewidth]{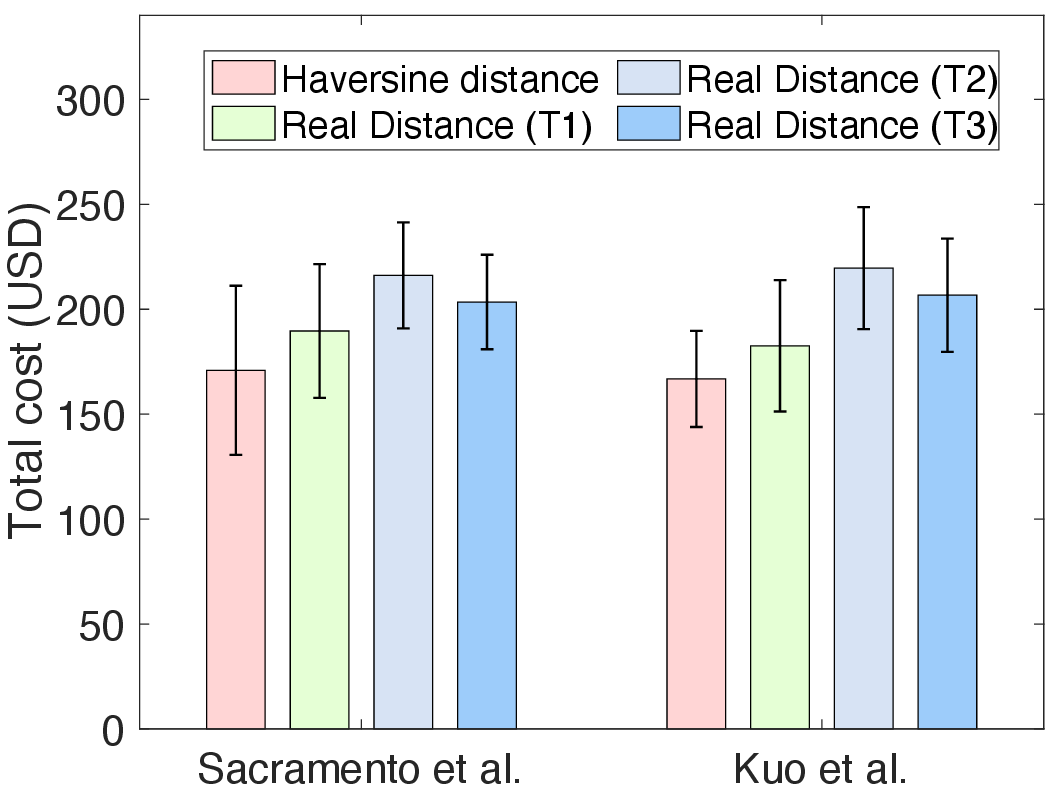}
		\caption {Effect of actual traveled distance and travel time.}
		\label{fig:motivation}
	\end{center}
	\vspace{-15pt}
\end{wrapfigure}

Existing VRP-D solutions face two limitations. Firstly, they rely on the simple Haversine distance in determining the optimal routes for trucks and drones. Secondly, existing solutions overlook the impact of fluctuating traffic conditions, which can significantly affect route efficiency. To better understand these limitations, we conducted a simulation study. In this study, we focused on how the basic distance calculation and varying traffic conditions influence the solution. For our experiment, we simulated a scenario with 50 customers. The locations of these customers were randomly selected, representing a variety of pickup and drop-off points within a specific date and time window. More detailed explanation of the simulation setup is presented in Section~\ref{sec:results}.

We implemented two popular VRP-D solutions: the approach by Sacramento \emph{et al.}~\cite{sacramento2019adaptive} and the method by Kuo \emph{et al.}~\cite{kuo2022vehicle}, to calculate the routes for trucks and drones. We then measure the total operational cost for both solutions. The results are presented in Fig.~\ref{fig:motivation}. At the same time, we also display the total operational costs calculated using actual travel times and distances, determined through the Google Distance Matrix API~\cite{googleapi} for comparison. Notably, we observed a significant discrepancy between the operational cost computed with the Haversine distance and actual travel distance and time denoted by $T1, T2$, and $T3$. Additionally, to investigate the impact of fluctuating traffic conditions, we measured the operational costs at different times. As shown in Fig.~\ref{fig:motivation}, the operational costs vary significantly each time. This variability underscores the dynamic nature of solution efficiency and the necessity for a novel VRP-D solution that incorporates a robust travel distance and time prediction module in computing the routes of trucks and drones.



\section{Design of VRPD-DT}
\label{sec:proposed_system}

\subsection{Overview}
\label{sec:overview}

Since VRP-D is in NP-Hard~\cite{schermer2019matheuristic}, we develop a heuristic solution. This section presents an overview of our heuristic solution based on the Variable Neighborhood Search (VNS) method~\cite{kuo2022vehicle}. The core strategy involves progressively seeking a local optimal solution, starting from a carefully chosen initial solution and systematically altering this solution through predefined neighborhood moves. Once a local optimum is identified, we implement a `shaking' procedure to escape from the local optimum and potentially discover a superior solution. Our approach's uniqueness lies in the seamless integration of a travel time prediction module, which plays a crucial role both in generating the initial solution and in the evaluation of subsequent solutions.

\begin{figure}[!htbp]
	\centering
	\includegraphics[width=.99\columnwidth]{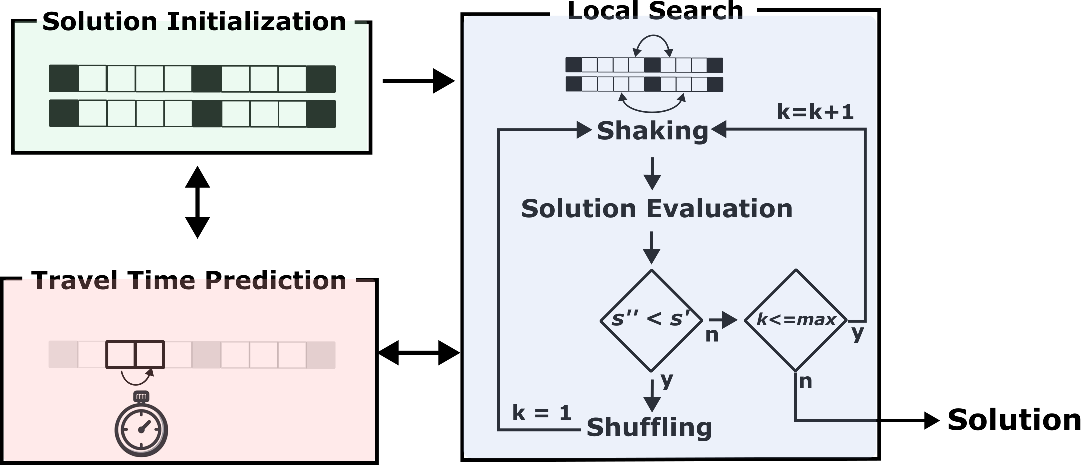}
	\caption {Overview of VRPD-DT.}
	\label{fig:overview}
\end{figure}

Fig.~\ref{fig:overview} shows the overall structure of our heuristic approach, which essentially comprises three modules: Solution Initialization, Travel Time Prediction, and Local Search. As illustrated in the figure, the process begins with the Solution Initialization Module, which creates a high-quality initial solution. This module relies on the Travel Time Prediction Module, utilizing a pre-trained machine learning model to estimate travel times and distances for an arc $(i,j) \in A$. Once the initial solution is created, it is then fed into the Local Search Module, starting the main iteration phase to determine the optimal solution.

The main iteration encompasses three steps: shaking, solution evaluation, and shuffling. The shaking procedure involves performing neighborhood moves to escape from a local optimum solution and search for a superior solution. The solution evaluation phase compares the newly generated solution $s''$ with the best solution $s'$ by utilizing the travel time prediction module. Specifically, solution evaluation is performed based on our new cost model that involves both the real-world traveled distance and estimated travel time computed using the travel time prediction model. Formally, the operational cost to traverse an arc $(i,j) \in A$ is calculated according to the cost model. 
\begin{equation}
\label{eq1}
	C_{ij} = (t_{ij}^{tr} \times c_{w}) + (d_{ij}^{tr} \times c_{veh}),
\end{equation}
\noindent where $C_{ij}$ denotes the total cost, $t_{ij}^{tr}$ is the estimated travel time, $c_{w}$ is for average wage rate, $d_{ij}^{tr}$ is the real-world traveled distance, and $c_{veh}$ denotes the vehicle operating cost. Consequently, if the new solution $s''$ fails to surpass the best solution $s'$, a shuffling operation is performed, and the main loop is repeated. Conversely, if the new solution is better than the best solution, the termination condition (\emph{i.e.,} the number of iterations $k$ is smaller than or equal to the threshold `$\mbox{max}$') is assessed. If the termination condition is satisfied, the solution is outputted as the final solution; if it is not, the main loop continues.

\subsection{Solution Representation}
\label{sec:solution_representation}

\begin{figure}[!htbp]
	\centering
	\includegraphics[width=.99\columnwidth]{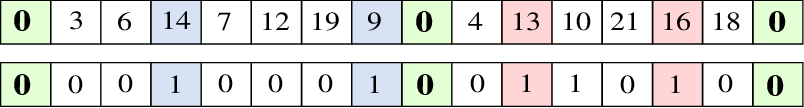}
	\caption {An example solution for VRPD-DT.}
	\label{fig:solution_representation}
\end{figure}

The solution of VRPD-DT is the routes of trucks and drones serving customers. It is basically sequences of serviced customers for trucks and drones. For the solution representation, we follow a similar approach for solution representation as outlined in the paper~\cite{kuo2022vehicle}. A solution comprises two vectors: the upper vector and the lower vector, as illustrated in Fig.~\ref{fig:solution_representation}. The upper vector delineates the routes of trucks and drones. In the upper vector, a non-negative integer denotes the customer id, and 0 signifies the commencement and termination of a route. The lower vector distinguishes whether a customer is served by a truck or a drone. Specifically, a 0 indicates that the customer is served by a truck. If the $i$-th element and the $i-1$-th element are both 1, then the customer signified by the $i$-th element is served by a truck. On the other hand, if the $i$-th element is 1 and the $i-1$-th element is 0, the customer identified by the $i$-th element is served by a drone. Lastly, if both the $i$-th elements in the upper and lower vectors are 0, it indicates that the drone and truck have returned to the depot. In particular, if $i$-th element is 0, and the $i-1$-th element is 1, it means that the drone is returned to the truck.

\begin{figure}[!htbp]
	\centering
	\includegraphics[width=.95\columnwidth]{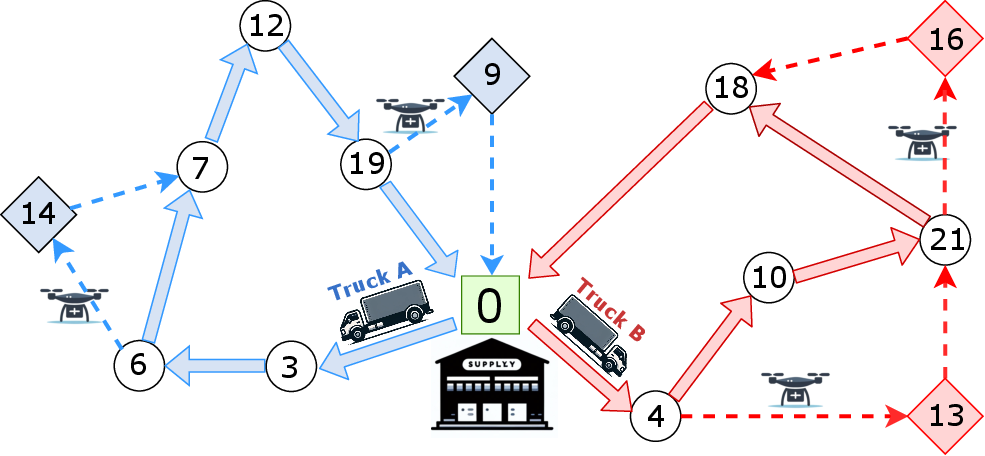}
	\caption {An illustration of the truck-drone operation corresponding to the solution shown in Fig.~\ref{fig:solution_representation}.}
	\label{fig:example_operation}
\end{figure}

The truck-drone operation corresponding to the solution shown in Fig.~\ref{fig:solution_representation} is illustrated in Fig.~\ref{fig:example_operation}. Here, the solid lines denote routes taken by trucks, while dotted lines represent those used by drones. As illustrated, two separate routes are established to serve 13 customer nodes. The truck-drone pairing in the first route begins at the depot (node 0), moving to node 3, and then to node 6. Given that the following entry in the lower vector is `1', the drone is dispatched from node 6 to serve node 14. Simultaneously, the truck progresses to node 7, rendezvousing with the returning drone, as indicated by the `0' in the lower vector. Post-retrieval of the drone, the truck-drone duo proceed to nodes 12 and 19, where the drone is once again deployed to serve node 9. Upon drone recollection, both units return to the depot to complete the route. The second route is executed simultaneously with the first route in a similar fashion.

\subsection{Solution Initialization}
\label{sec:solution_initialization}

The solution initialization module is designed to generate a high-quality initial solution which is provided as input to the local search module. More specifically, given the number of trucks, the initial solution is generated as follows. First, we simplify the original problem, treating it as a vehicle routing problem (VRP) that considers trucks only. This simplified problem is then addressed using a straightforward nearest neighbor approach. A salient aspect is that in computing the nearest neighbor, we use the cost (Eq.~\ref{eq1}) built based on the travel time prediction module. This means that starting from the depot, the truck consistently moves to the closest node with the minimum cost, provided that the constraints related to capacity and operating time limit are met. The solution generated via this nearest neighbor strategy is further refined using the 2-OPT approach, followed by a further improvement using the Findsortie algorithm~\cite{sacramento2019adaptive}.

\subsection{Travel Time Prediction}
\label{sec:travel_time_prediction}

The travel time prediction module is designed for estimating the travel time $t_{ij}^{tr}$ for an arc $(i,j) \in A$, which is a key component in calculating the estimated cost $C_{ij}$ at any given time. This module is particularly important in accounting for the variability of travel times throughout the day. It takes the pick-up and drop-off GPS locations, along with the date and time of pick-up, as input to predict trip duration. The prediction is based on historical trip data and relevant features, following the methodology of an existing ML-based travel time prediction module as detailed in~\cite{khaled2022gsta}. These features include a comprehensive set of parameters that are instrumental in providing as precise a cost estimation as possible:

\begin{itemize}
	\item \textbf{Location Features}: pickup longitude, pickup latitude, dropoff longitude, dropoff latitude.
	
	\item \textbf{Date/Time Features}: day of week, day of month, hour, weekend, work day, peak hour, public holiday, trip duration.
	
	\item \textbf{Weather Features}: temperature, dew, humid, rain, snow, visible, fog, thunder, tornado, clear, haze, heavy rain, heavy snow, light rain, light snow.
	
	\item \textbf{Other Features}: average vehicle speed, trip miles. 
\end{itemize}

The travel time prediction module plays a pivotal role in our heuristic solution, both in generating effective initial solutions and in evaluating them. The travel time prediction module is trained on the comprehensive New York City taxi dataset~\cite{NYC-tlc}, which encompasses over 21.9 million taxi trips and includes a diverse array of features. Consequently, our simulation was specifically conducted in the New York City area. It is important to note that to adapt our solution for last-mile delivery in different regions, a new travel time prediction model must be trained using data relevant to the target area.

\begin{figure}[!htbp]
	\centering
\includegraphics[width=.8\columnwidth]{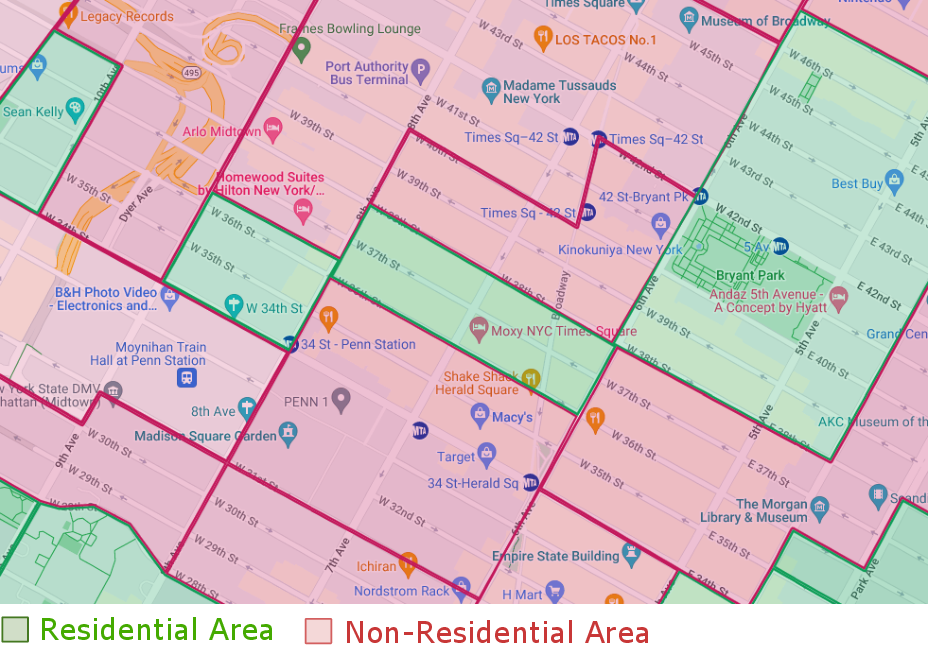}
\caption {An example of residential areas for the NYC taxi dataset.}
\label{fig:residential}
\end{figure}


Our travel time prediction module is specifically tailored to meet the unique requirements of the last-mile delivery application, distinguishing it from generic models such as~\cite{khaled2022gsta}. Key to our approach is the development of a residential area-aware model. We capitalize on the observation that last-mile delivery trucks primarily traverse residential areas with typically sparse traffic, \emph{i.e.,} minimal traffic fluctuations. This insight allows us to streamline model training by focusing on estimating travel times primarily for routes involving non-residential roads (\emph{e.g.,} See Fig.~\ref{fig:residential}), significantly reducing computational costs. For such routes, our model estimates travel times, while for others, it simplifies the calculation to distance traveled divided by average vehicle speed. Moreover, we refined our data pre-processing phase, excluding trips exceeding a certain distance threshold. This adjustment, recognizing that customer-to-customer travel is generally short, reduces our dataset to 18.6 million trips, enhancing the model's relevance to last-mile delivery scenarios.

\subsection{Local Search}
\label{sec:neighborhood_moves}

The local search module is designed to find near-optimal solutions. Tasking the initial solution as input, it executes in three distinct phases: shaking, solution evaluation, and shuffling. The shaking phase is used to escape away from the local minimum solution, thereby enabling a more efficient exploration of the solution space and the generation of high-quality solutions. This shaking process employs eight standardized neighborhood move procedures: (i) Random swap node, (ii) Random swap whole, (iii) Random insertion node, (iv) Random insertion whole, (v) Random reverse node, (vi) Random reverse whole, (vii) Remove sortie node, and (viii) Add sortie node. Random swap swaps nodes in both the upper and lower vectors, while random insertion inserts a node at a specific position within the vectors. Random reverse modifies the neighborhood by completely reversing a selected section of the vectors. The `whole' moves differ from the regular moves as they alter both the upper and lower vectors simultaneously, resulting in a more significant modification. The last two moves, \emph{i.e.,} remove and add sortie, only change the lower vector thereby modifying the transportation mode (drone to truck and vice-versa). Specifically, the remove sortie action excludes a certain node from being part of a sortie by changing the corresponding value in the lower vector from `1' to `0'. This effectively establishes a new rendezvous point for the truck and drone. Conversely, the add sortie operation attempts to integrate a certain node into a sortie by adjusting the lower vector value from `0' to `1'. However, this does not guarantee that the node will be included in a drone sortie because a `1' in the lower vector signifies that the node could be serviced by either a drone or a truck, depending on which is available.

After the shaking phase, the solution evaluation phase is executed. In this phase, the quality of the generated solution is evaluated according to the penalized cost function as shown below. 

\begin{align*}
	p_z
	= &Z +  \sum_{v \in T \cup D}p \cdot max(0, \sum_{i,k \in r_v^t}\sum_{j \in r_v^d}y_{ijk}^v(t_{ij}'+t_{jk}')-E)   \\
	&+ \sum_{v \in T}p \cdot max(0,\sum_{i \in r_v^t}q_i - Qt) \\
	&+ \sum_{v \in D}p \cdot max(0,\sum_{i \in r_v^d}q_i - Qd) \\
	&+ \sum_{v \in T \cup D}p \cdot max(0,a_{n+1}^v - T_{max}, a_{n+1}^{'v} - T_{max}) \\
	&+ \sum_{v \in T \cup D}p W(v),
\end{align*}

\noindent where $Z$ is the total operational cost which is defined as follows. 

\begin{displaymath}
	Z = \sum_{v \in T \cup D}(\sum_{i \in C} \sum_{j \in C} C_{ij}x_{ij}^v + \sum_{i \in C} \sum_{j \in C} \sum_{k \in C}(C'_{ij}+C'_{jk})y_{ijk}^v),
\end{displaymath}

\noindent where $x_{ij}^v$ is 1 if the arc $(i,j) \in A$ is used by the truck for delivery. The terms following $Z$ define the penalizing costs, \emph{i.e.,} the term $max(0, \sum_{i,k \in r_v^t}\sum_{j \in r_v^d}y_{ijk}^v(t_{ij}'+t_{jk}')-E)$ is the endurance penalty, ensuring that the total drone operation time remains less than $E$; $max(0,\sum_{i \in r_v^t}q_i - Qt)$ denotes the truck load penalty, indicating that a truck can only deliver parcels to a customer whose demand is less than its load capacity; $max(0,\sum_{i \in r_v^d}q_i - Qd)$ represents the drone load penalty; $max(0,a_{n+1}^v - T_{max}, a_{n+1}^{'v} - T_{max})$ is the duration penalty, meaning that all deliveries must be accomplished within $T_{max}$; finally, $\sum_{v \in U}p W(v)$ is the penalizing cost incurred when the arrival time of truck $v$ (or drone), \emph{i.e.,} $a_i^v$ (or ${a'}_i^v$) exceeds the latest allowable service time $e_i$ for node $i$.

After the solution evaluation phase, if the new solution is not as good as the best one found so far, the algorithm decides to explore the search space more thoroughly. To achieve this, it performs the shuffling operation, which rearranges the neighborhood list to improve the effectiveness of the search space exploration.

\section{Simulation Results}
\label{sec:results}

In this section, the performance of our approach is evaluated. We implemented and executed our solution and compared the performance with a state-of-the-art VRPTWD heuristic algorithm~\cite{kuo2022vehicle}. The experiments were carried out on a PC equipped with an AMD Ryzen 7 7840HS CPU, 16GB RAM, and NVIDIA RTX 4050 GPU, operating on Windows 11. All codes were developed in Python 3.10.

\subsection{Test Instances}
\label{sec:test_instances}

To evaluate the performance of our approach, we created a novel dataset modeled after the one used by~\cite{kuo2022vehicle}. However, unlike their simple grid map approach, our dataset uniquely incorporates real-world locations from New York City~\cite{NYC-tlc}, offering a more complex and realistic testing environment. We conducted performance evaluations using various numbers of customer nodes, with a maximum of 50. These nodes were randomly distributed throughout the New York City area, with the depot strategically placed at a central location.

To enable direct performance comparison with the SOA method~\cite{kuo2022vehicle}, we assigned the time window for each customer node as specified in their paper~\cite{kuo2022vehicle}. More specifically, two parameters, namely the time window density (\%TW) and time window width ($w$) are randomly selected in the range 25\% to 100\%, and 30 to 120 minutes, respectively. Here \%TW is the percentage of customer nodes with time windows, and $w$ is the interval of the time window. For example, for each test instance, a particular fraction of customer nodes (\%TW) are assigned with time windows with a random time window width ($w$). More detailed information about the random time window generation method can be found in~\cite{kuo2022vehicle}.

\subsection{Key Metric}
\label{sec:metrics}

Our solution aims to enhance the accuracy of operational cost estimation under dynamic traffic conditions. To evaluate the effectiveness, we measured how closely the operational costs calculated using our solution aligned with the actual costs, compared with the SOA method~\cite{kuo2022vehicle}. More specifically, we first determine the operational cost using the SOA method, denoted as $C_{VRPD}$, and then using our method, denoted as $C_{VRPD-DT}$. The accuracy of each method is quantified by the percentage discrepancy from the actual cost $C_{ACTUAL}$, calculated as $\frac{|C_{VRPD} -  C_{ACTUAL}|}{C_{ACTUAL}}$ for the SOA method and $\frac{|C_{VRPD-DT} -  C_{ACTUAL}|}{C_{ACTUAL}}$ for our method, respectively. The actual cost $C_{ACTUAL}$ is computed using the Google Maps API to measure the travel distance and the travel time for each arc on the optimal truck routes, taking into account the start time of journey for each arc.

\subsection{Cost Estimation Accuracy}
\label{sec:cost_accuracy}

\begin{figure}[!htbp]
	\centering
		\includegraphics[width=.8\columnwidth]{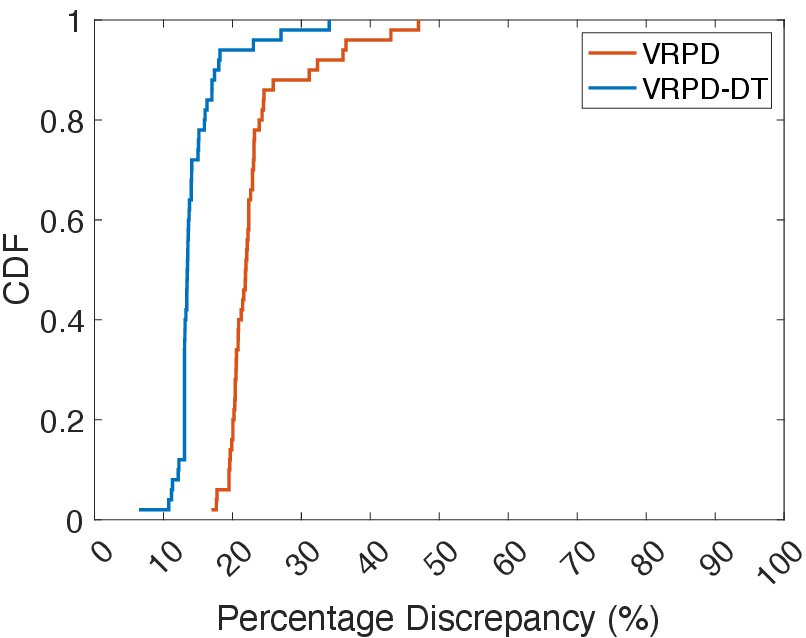}
	\caption {The cost estimation accuracy for the SOA and our approaches.}
	\label{fig:impact_of_customers}
\end{figure}


To assess the accuracy of solutions obtained with both the SOA method and our approach, we conducted experiments across various delivery scenarios with 50 randomly selected customer locations. These experiments were conducted 50 times to ensure robustness. The outcomes of these tests are presented in the form of a cumulative distribution function (CDF) graph, as shown in Fig.~\ref{fig:impact_of_customers}. The findings show that our method reduced the average and maximum percentage discrepancies by 37.6\% and 27.6\%, respectively, compared with the SOA approach. The results underscore the significant impact of the travel time prediction of VRPD-DT in obtaining more accurate solutions.

\subsection{Effect of Number of Customers}
\label{sec:effect_of_num_of_neighbors}

\begin{figure}[!htbp]
	\centering
	\includegraphics[width=.8\columnwidth]{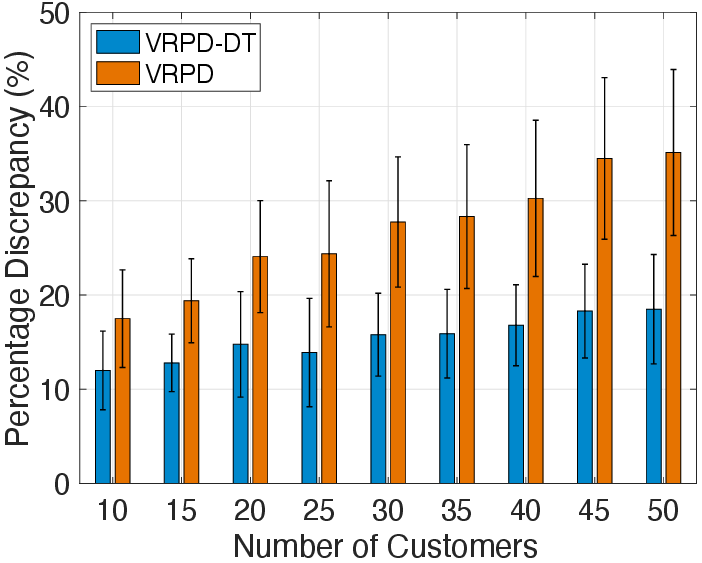}
	\caption {Effect of number of customers.}
	\label{fig:impact_of_customers}
\end{figure}


In this section, we address an important question: how does the number of customers affect the performance of our approach and the SOA method? To answer this question, we analyzed the percentage discrepancy for both methods while varying the customer count. Figure~\ref{fig:impact_of_customers} illustrates our findings. Notably, there is a modest rise in the discrepancy for our method as customer numbers grow, due to the increased number of arcs to reach more customers, leading to higher cumulative errors. However, an interesting observation was that the SOA method exhibits a markedly higher growth rate in the percentage discrepancy compared to our solution. Specifically, the SOA method's discrepancy surged by 100.7\% when customer numbers rose from 10 to 50, whereas our method saw a smaller increase of 54.2\%.

\subsection{Ablation Study}
\label{sec:ablation_study}

\begin{figure}[!htbp]
	\centering
	\includegraphics[width=.8\columnwidth]{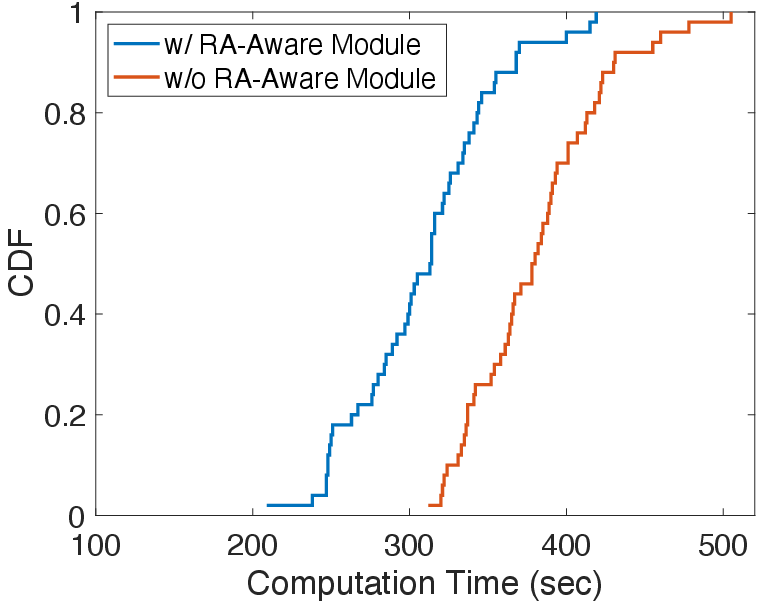}
	\caption {Effect of the residential area (RA)-aware travel time prediction model on the computational time.}
	\label{fig:computation_time}
\end{figure}


We performed an ablation study to assess the impact of incorporating the residential area (RA)-aware travel time prediction module, focusing particularly on its efficiency in reducing the solution computation time. Reducing the solution computation time is crucial, unlike reducing the time for the one-time model training process, since solution computation is a recurring task. For consistency, we set the number of randomly selected neighbors at 50 in each trial. We conducted repeated experiments to measure the computation times both with and without the RA-aware travel time prediction module. The results, illustrated in Figure~\ref{fig:computation_time}, reveal a significant reduction in solution computation time with the integration of the residential area-aware travel time prediction module, achieving a decrease of 18.8\%, compared to the generic travel time prediction model. 

\begin{figure}[!htbp]
	\centering
	\includegraphics[width=.8\columnwidth]{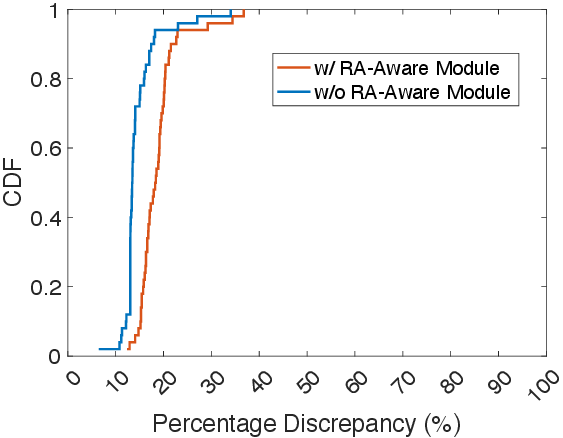}
	\caption {Effect of the residential area (RA)-aware travel time prediction model on the solution accuracy.}
	\label{fig:ablation_accuracy}
\end{figure}


An anticipated limitation of the RA-aware travel time prediction module is, however, a potential reduction in the accuracy of solutions. The reason for the degraded accuracy can be attributed to the fact that travel time predictions are not performed for arcs $(i, j) \in A$ for which both customer locations $i \in C \cup D$ and $j \in C \cup D$ are in residential areas, unlike the generic model that performs predictions for all arcs. To quantify this potential accuracy trade-off, we compared the accuracy of solutions with and without the RA-aware module. As shown in Figure~\ref{fig:ablation_accuracy}, we observed an average accuracy decrease of 4.9\% with the RA-aware module. This suggests that while the RA-aware module can significantly reduce the solution computation time, such benefits come with a slight compromise in the accuracy of solutions.

\section{Conclusion}
\label{sec:conclusion}

We have presented a novel problem called the vehicle routing problem with drones under dynamically changing traffic conditions (VRPD-DT) to address the limitation of existing VRP-D solutions that produce sub-optimal solutions as the solution can be potentially altered as traffic conditions change. VRPD-DT aims to minimize the total operational cost of trucks and drones that work in tandem to deliver parcels to customers under dynamically changing traffic conditions. A new cost model is created to represent the operational cost more accurately by taking into account the real-world traveled distance and the actual travel time in time-varying traffic conditions through incorporation of a machine learning-based travel time prediction. A variable neighborhood descent (VND)-based heuristic algorithm integrated with the new cost model is designed to solve VRPD-DT. A simulation study was performed to demonstrate the effectiveness of our approach in comparison with a SOA VRPTWD algorithm. The results indicated that our algorithm outperforms under various delivery scenarios. Our future work is to enhance our proposed approach by integrating real-world drone-specific factors. This includes accommodating drones with varying ranges, accounting for wind resistance, monitoring drone battery levels, and considering the sizes and weights of packages that drones carry.

\appendix





\bibliographystyle{named}
\bibliography{ijcai24}

\end{document}